\documentclass{icrc}

\usepackage{times}
\usepackage{graphicx}

\begin{document}

\title{EAS data at the mountain level and
a shape of the CR spectrum beyond the break.}
\author[1]{S. B. Shaulov}
\affil[1]{P.N. Lebedev Institute, Leninsky prospect 53,
117924 Moscow, Russia}

\correspondence{shaul@sci.lebedev.ru}

\runninghead{S. B. Shaulov: CR spectrum }
\firstpage{1}
\pubyear{2001}

\maketitle

\begin{abstract}
In the most works which deal with EAS the CR energy spectrum is
deduced by means of the model defined dependence
$E_0=a\cdot{N_e}^{\alpha}$. An electron total number $N_e$ is
evaluated by integration of the NKG-function f(r). This algorithm
breaks down for young EAS with age parameter $s\sim0$. This work
shows, that part of the young EAS becomes large in the range
$N_e\geq10^7$, it cause to divergency of the $N_e$ integral for
them and distorts the shape of EAS (CR) spectrum.  A final analysis
of the experimental data permits to conclude that EAS spectrum has
local maximum at $N_e\sim10^9$, which results in a decrease of the
EAS spectrum slope for $N_e\geq10^7$ (inverse break or ``knee''). A
local maximum can arise because of the additional CR component in
the range $E_0\geq10$ PeV.
\end{abstract}

\section{Introduction}

A total electron number $N_e$ spectrum is analyzed for EAS (extensive
air showers), detected at Tien Shan altitude (685 {\rm g/cm}$^2$) in
experiment ``Hadron'' \citep{adr86}.  The range under study
$N_e=10^6-10^9$ involves EAS spectra break at $N_e\simeq1.5\cdot10^6$
and the main aim of this work is to investigate the spectrum shape
beyond the break.

A peculiarity in EAS spectrum is often named as ``knee'', implying
that an inverse changing of EAS spectrum slope is observed beyond
the break. A value of this changing is differ in the dissimilar
experiments, i.e. depends of the detecting level, installation
design, in particular the kind of detectors and them separation,
and methods of $N_e$ estimation.

It will be shown that a peculiarity vary in shape to comparison with
one, received in \citep{nest95}, if $N_e$ dependence of the EAS age
parameter s is taken into account. A considerable increase of the
young EAS part with a small s can decreases the EAS intensity for a
large $N_e$, that cause to maximum formation in EAS spectrum at
$N_e\sim10^8$.

\section{Experimental results and methodic.}

The total electron number $N_e$ as usual is determined by integral:
\begin{equation}
N_e=2\pi\int\limits_0^{\infty}{f_{NKG}(r,\hat{s})rdr},
\label{fl:1}
\end{equation}
where r is a distance from EAS axis and
\begin{equation}
f_{NKG}(r)=C(s)\cdot(\frac{r}{r_m})^{s-2.0}\cdot(1-\frac{r}{r_m})^{s-4.5},
\label{fl:2}
\end{equation}
is NKG (Nishimura-Kamata-Greysen) functions, $r_m$ - Molier radius
and a normalized coefficient
\begin{equation}
C(s)=\frac{1}{2\pi}\frac{\Gamma(4.5-s)}{\Gamma(s)\Gamma(4.5-2s)}
\simeq 0.366s^2(2.07-s)^{1.25}.
\label{fl:3}
\end{equation}
LDF (lateral distribution function) is defined by the follow
expression:
\begin{equation}
\rho_{e^-}=\frac{N_e}{r_m^2}f_{NKG}
\label{fl:4}
\end{equation}
In a frame of this approximation the values of the individual
NKG-function are determined by an estimation of the parameter
$s=\hat{s}$ (LDF slope) and a normalization to a sum of the
experimental $e^{-}$ densities $\rho_i$ in the scintillator
detectors.

An algorithm for $N_e$ estimation cited above is true for EAS with
s-parameter $\sim1$ or more, but it breaks down at $s\rightarrow0$,
as $f_{NKG}\sim{1/r^2}$ in that case and an integral in Eq.
(\ref{fl:1}) is diverged. A pole in $s=0$ isn't a physical one and
to get rid of the divergency one must introduce a cut-off for the
$f_{NKG}(r)$ in area close to $r\sim0$. $N_e$ is defined by the
numerical integration in that case \citep{sha96}. A
cut-off radius is selected of order $r_0\sim0.5-1~$m.

As usual EAS part with $s\sim0$ is negligible but it increases with
$N_e$ in particular beyond the break for $N_e\geq5\cdot10^6$
\citep{cher99}.

\begin{figure}[t]
\includegraphics[width=8.3cm]{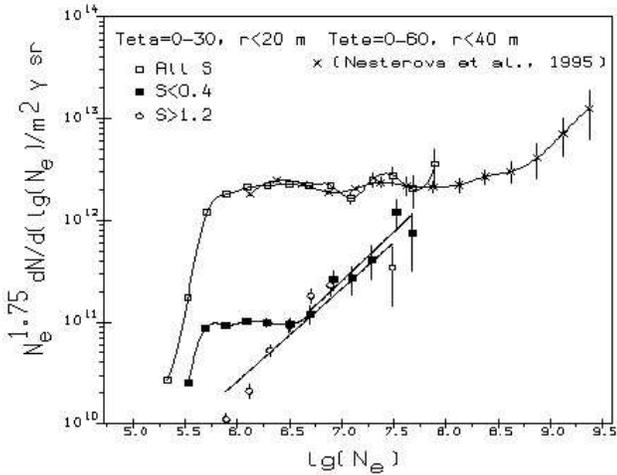}
\caption{$N_e$ spectra, multiplied by $N_e^{1.75}$, for all EAS,
young EAS with small s-parameter $s<0.4$ and old EAS with $s>1.2$.
EAS spectrum from \protect{\citep{nest95}} is
marked by skewed crosses.}
\label{f:1}
\end{figure}

An all EAS spectrum is compared with one's  for the most young EAS
with $s<0.4$  and the most old with $s>1.2$.  The spectra are
multiplied by $N_e^{1.75}$, At the left of Figure everything spectra
are limited by the EAS detecting threshold.

A s-parameter distribution is changed above the break, as EAS part
with $s<0.4$ increases from $\sim5\%$ for $N_e\sim3\cdot10^6$ up to
$\sim40\%$ for $N_e\sim3\cdot10^7$. At the same range the part of
old EAS is increased too and it is worthy to empathize that them
part for small $N_e$ is negligible. The last circumstance
explaines an increase of the s-parameter average value $\overline{s}$
for $N_e>10^7$ \citep{cher99}.

A EAS spectrum from \citep{nest95} is shown in Fig. \ref{f:1} too. It
was received with much more EAS statistic by means of increase the
selection radius from $r\leq20$ m up to $r\leq40$ m and range zenith
angles up to $\theta\leq60^{\circ}$.  Attention is drawn to the fact
that an EAS intensity increase for $N_e>10^8$ may be connected with
young EAS to a large degree.

\begin{figure}[t]
\includegraphics[width=8.3cm]{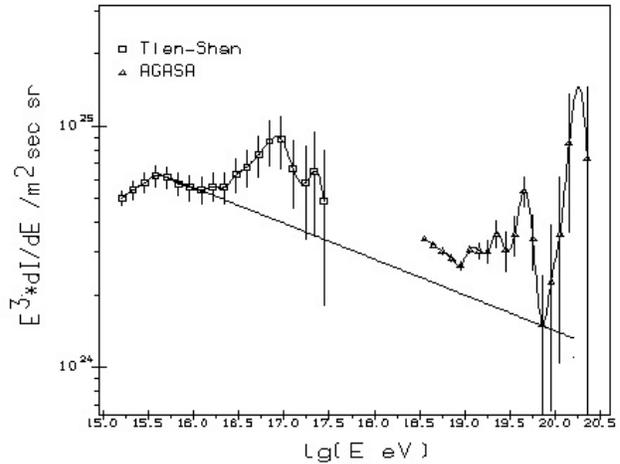}
\caption{CR energy spectrum, derived from Tien Shan and AGASA EAS
data. Spectrum is multiplied by $E^3$.}
\label{f:2}
\end{figure}

NKG cut-off in range $r<1$ {\rm m} for EAS with $s<0.4$ decreases the
average value $N_e$ and changes the spectrum shape. The resulting CR
energy spectrum is shown in Fig. \ref{f:2} in comparison with AGASA
one \citep{takeda98}. A conversion from $N_e$ to energy was made in
response to quasi-scaling model \citep{erl_diss} by means of
expression $E=15.1\cdot{N_e^{0.84}}$ {\rm PeV}.

NKG limitation near the EAS axis decreases $N_e$ and causes to
maximum formation at energy $E\sim5\cdot10^{16}$ {\rm eV}. In
addition the intensities of the two spectra (Tien Shan and AGASA)
were brought into coincidence in the intermediate area. Furthermore a
resemblance between these parts of CR spectra in the ranges
$E=3\cdot10^{15}-3\cdot10^{17}$ {\rm eV} and
$E=3\cdot10^{18}-3\cdot10^{20}$ {\rm eV} can discussed.

\section{Discussion}

A true reconstruction of the original shape for CR spectrum is of
the utmost significance for the ``knee'' interpretation.
Initially the ``knee'' model was based on two assumptions: i) a
break in the CR nuclear spectra occurs at a value of the magnetic
rigidity $R\simeq3-4$ {\rm PV}, ii) protons are the dominant CR
component ($\sim40\%$) up to energies in few PeV. A very simple
``knee'' explanation results in this case. A break in CR spectrum
is connected with one in protons component and ``knee'' is formed
by more heavier nuclei.

This model breaks down if the experimental results, which were
received in succeeding years, are taken into account.

The direct measurements \citep{grig70,burnett83,burnett95} have
shown that protons part isn't exceed $25\%$ even at energies in
hundreds TeV and continues to decrease up to the break in accordance
with EAS data \citep{chatelet91}. It was
confirmed by investigation of the $N_e$ spectrum for EAS with
$\gamma-$families in experiment ``Hadron'', in which a small value of
the magnetic rigidity for break in nuclear spectra was received
$R\simeq0.1$ {\rm PV} \citep{sha99}.

A model with break for $R\simeq0.1$ {\rm PV} only can't be true,
because it gives too heavy CR composition for high energies. It seams
that a situation is saved by another experiment ``Hadron'' result
about arising of the additional component in CR, consisting mainly
from protons, beyond the CR spectrum break \citep{sha01}.  The data
of this work about EAS spectrum shape agree well with such
conclusion.

It means that transformed ``knee'' model consists from
break in nuclear spectra for the magnetic rigidity  $R\simeq0.1$
{\rm PV} and a contribution of the additional CR component in the
radiation for energies $E\geq5-10$ PeV.

The absence of this maximum or them more small value in CR spectrum
($E\sim5\cdot10^{16}$ eV) for the measurements at sea level may be
explained by the large separation between detectors in the
installation with large square, that can cause to the loss of the
most young EAS, or the difference of this component absorption in
the atmosphere in comparison with nuclei one's.

\begin{acknowledgements}
I am thanks all experiment ``Hadron'' participants, which
heavy-duty operation permits to receive these experimental data.
\end{acknowledgements}


\begin{thebibliography}{99}

\bibitem[Abdrashitov et al., 1986]{adr86}  Abdrashitov S. F. et al.,
Installation ``Hadron'' for an investigation of the primary CR and
characteristics of the nuclear interactions in the atmosphere by
EAS, EC and Cherencov light methods, Izv. AN USSR, ser. fiz., 1986,
v. 50, N11, p. 2203-2207

\bibitem[Nesterova et al., 1995]{nest95}  Nesterova N.M., Chubenko
A.P., Djatlov P.A. ,  Vildanova L.I., The primary cosmic ray spectrum
at $2\cdot10^{13}-2\cdot10^{18}$ eV and its peculiarity above
$10^{18}$ eV according to Tien-Shan data, in Proc. 24th ICRC,
Roma, Italy, v. 2 (1995) 748

\bibitem[Shaulov, 1996]{sha96}  Shaulov S. B., A method of the EAS
characteristics determination in the EAS associated with
$\gamma-$families and scaling violation, Preprint FIAN N60, Moscow,
1996, p. 3-37

\balance/

\bibitem[Cherdyntceva et al., 1999]{cher99}  Cherdyntceva K. V.,
Chubenko A. P.,    Nesterova N.  M.,  Pavluchenko V. P.,
Shaulov S. B., A changing of the $\gamma$-family and EAS in range
$N_e=3\cdot10^6-10^7$, Preprint FIAN N7, 1999, 3-15

\bibitem[Takeda et al., 1998]{takeda98}  Takeda M., Hayashida N.,
Honda K. et al., Extension of the cosmic-ray energy spectrum beyond
the predicted Greizen-Zatsepin-Kuz'min cutoff, Phys. Rev. Letters,
vol.  81, N6, 1998, pp. 1163-1166

\bibitem[Erlykin, 1986]{erl_diss}  Erlykin A. D., A many-dimensional
analysis of the hadron cascades in atmosphere for nuclear and
astrophysical CR investigations, Doctor diss., FIAN, Moscow, 1986

\bibitem[Grigorov et al., 1970]{grig70} Grigorov N. L., Nesterov V.
E., Rapoport I. D. et al., Investigation of the PCR high and
superhigh energy spectrum at cosmic stations ``Proton'', Nuclear
Physics, 1970, II, N5, p. 1058-1069

\bibitem[Burnett et al., 1983]{burnett83}  Burnett T.H.,  Dake S.,
Fuki M.  et al., Proton and Helium energy spectra above 1 TeV for
primary cosmic rays, Phys. Rev. Lett.,{\bf 51}, N11 (1983) 1010-1013;

\bibitem[Burnett et al., 1995]{burnett95} JACEE collaboration, Energy
spectra and elemental composition  of nuclei above 100 TeV from a
series of the JACEE balloon flight, in Proc.  24th ICRC, Roma, Italy,
v.2, 1995, p. 707

\bibitem[Chatelet et al., 1991]{chatelet91}  Chatelet E.,
Danilova T.V.,   Erlykin A.D.,   Pavluchenko V.P.,  Procureur J.,
Re-analysis of the primary mass composition in the ``knee region'' on
the basis of the Tien-Shan data and the independent-nucleon model of
A-A interactions, J.Phys.  G:Nucl. Part. Phys., {\bf
17}(1991)1427-1439

\bibitem[Shaulov, 1999]{sha99}  Shaulov S. B., CR composition in area
of ``knee'' and a contribution of the single, close source, Preprint
FIAN N8, Moscow 1999

\bibitem[Shaulov, 2001]{sha01} Shaulov S. B., Experimental evidences
of two component model for CR composition around the ``knee'', this
conference, 2001

\end{thebibliography}
\end{document}